\documentclass[12pt,preprint]{aastex}
\usepackage{amsmath}

\slugcomment{submitted to ApJ main journal}
\shorttitle{Magnetic Field in the M~87 Jet}
\shortauthors{Stawarz et al.}

\begin{document}

\title{On the Magnetic Field\\
in the Kiloparsec-Scale Jet of Radio Galaxy M~87}

\author{\L ukasz Stawarz$^{1, 2}$, Aneta Siemiginowska$^{1}$,\\ Micha\l \, Ostrowski$^{2}$ and Marek Sikora$^{3}$}
\affil{$^1$Harvard-Smithsonian Center for Astrophysics,\\ 60 Garden Street, Cambridge, MA 02138, USA}
\affil{$^2$Astronomical Observatory, Jagiellonian University,\\ ul. Orla 171, 30-244 Krak\'ow, Poland}
\affil{$^3$Nicolaus Copernicus Astronomical Center,\\ ul. Bartycka 18, 00-716 Warszawa, Poland}

\altaffiltext{}{stawarz@oa.uj.edu.pl ; asiemiginowska@cfa.harvard.edu ; mio@oa.uj.edu.pl ; sikora@camk.edu.pl}

\begin{abstract}

Several low-power kiloparsec-scale jets in nearby radio galaxies are known for their synchrotron radiation extending up to optical and X-ray photon energies. Here we comment on high-energy $\gamma$-ray emission of one particular object of this kind, i.e. the kiloparsec-scale jet of M~87 radio galaxy, resulting from comptonization of the starlight photon field of the host galaxy by the synchrotron-emitting jet electrons. In the analysis, we include relativistic bulk velocity of the jet, as well as the Klein-Nishina effects. We show, that upper limits to the kiloparsec-scale jet inverse-Compton radiation imposed by {\it HESS} and {\it HEGRA} Cherenkov Telescopes --- which detected a variable source of VHE $\gamma$-ray emission within 0.1 deg ($\sim 30$ kpc) of the M~87 central region --- give us an important constraint on the magnetic field strength in this object, namely that the magnetic field cannot be smaller than the equipartition value (referring solely to the radiating electrons) in the brightest knot of the jet, and most likely, is even stronger. In this context, we point out a need for the amplification of the magnetic energy flux along the M~87 jet from the sub-parsec to kiloparsec scales, suggesting the turbulent dynamo as a plausible process responsible for the aforementioned amplification.

\end{abstract}

\keywords{galaxies:  jets --- galaxies: individual(M~87) --- magnetic fields --- radiation mechanism: non-thermal}

\section{Introduction}

The pure non-thermal nature of the multiwavelength emission of extragalactic jets results in a fact, that many of their parameters are basically unknown. Intensity of the jet magnetic field is the exemplary unknown in all jet models. The situation is even less clear on large scales ($\geq 1$ kpc) than on the small (sub-pc and pc) scales, as typically the observed spectrum of large-scale jets consists of the synchrotron emission alone, without synchrotron self-absorption features or the inverse-Compton component. Therefore, the usual approach is to assume energy equipartition between the large-scale jet magnetic field and synchrotron radiating electrons, obtaining thus $B_{\rm eq} \sim 10^{-6} - 10^{-3}$ G. The standard justification for the equipartition assumption is that the inverse-Compton X-ray emission detected from a number of hot-spots and lobes in powerful radio sources often (although not always) yields $B \approx B_{\rm eq}$ \citep[see, e.g.,][and references therein]{kat04}. This, however, cannot be really taken as a proof for the magnetic field--radiating electrons energy equipartition in the case of the jet flows, as the physical processes responsible for the evolution of radiating particles and magnetic field within terminal shocks and extended lobes can differ substantially from the respective processes that take place within large-scale jets themselves \citep[see, e.g., a discussion on the magnetic field structure within the hot-spots and lobes by][]{blu00}.

Unfortunately, the X-ray emission detected recently from a number of large-scale quasar jets \citep[e.g.,][]{sch00,sie02,sie03,sam04} cannot give us a definite answer on the magnetic field intensity in these objects. First, it is not well established if this X-ray emission is synchrotron or inverse-Compton in origin \citep[see a discussion in][]{sta05}. Second, poorly constrained relativistic bulk velocities of the large-scale jets influence significantly the inferred values of the jet parameters in both cases. As a result, one can only say that if the X-ray emission of large-scale quasar jets is indeed due to the inverse-Compton scattering of the cosmic microwave background radiation \citep{tav00}, then it is possible to find a value of the jet Doppler factor that allows energy equipartition between the magnetic field and the radiating electrons \citep[or even equipartition between magnetic field energy and the total particle's bulk energy,][]{ghi01} in a certain object. This, however, does not mean that the energy equipartition is fulfilled. In fact, \citet{kat04} argued that in a framework of the inverse-Compton hypothesis the more plausible interpretation leads to sub-equipartition magnetic field within the large-scale quasar jets. We note that the analysis of jet dynamics suggests that the powerful quasar jets are most likely matter-dominated, at least on the large scales \citep[see, e.g.,][and references therein]{sik04}. Still, other models involving Poynting flux dominated outflows \citep{bla02} cannot be rejected with no doubts.

Contrary to the large-scale quasar jets, the synchrotron origin of the X-ray emission of the kpc-scale flows in low-luminosity radio galaxies is well established \citep[see][and references therein]{sta05}. In addition, the two-sideness of the FR I radio structures suggests much lower bulk velocities of these jets as compared to their powerful quasar-hosted analogues. All of these constraints give the unique opportunity to estimate more accurately magnetic field intensity within some nearby FR I jets by studying their inevitable inverse-Compton $\gamma$-ray emission. Unfortunately, due to insufficient sensitivity of the present $\gamma$-ray detectors, such analysis can be performed only for the closest FR I sources.

\citet{sta03} considered high energy (VHE) $\gamma$-ray emission produced by the kpc-scale jets in nearby low-power radio galaxies of the FR I type. Optical and X-ray emission detected recently from a number of such objects indicate that these jets are still relativistic on the kpc scale and that they contain ultrarelativistic electrons with energies up to $100$ TeV \citep[see a discussion in][]{sta05}. Therefore, some of the nearby FR I jets can be in principle VHE $\gamma$-ray emitters due to the inverse-Compton (IC) scattering off ambient photon fields which, at kpc distances from the active nuclei, are expected to be still relatively high. For example, following \citet{tsa95} it can be found that the bolometric energy density of the stellar emission at $1$ kpc from the center of a typical luminous elliptical galaxy is on average $U_{\rm star, \, bol} \approx 10^{-9}$ erg cm$^{-3}$ (which can be compared with the local value of the cosmic microwave background radiation, $U_{\rm cmb} = 4 \times 10^{-13}$ erg cm$^{-3}$). In the particular case of radio galaxy M~87, \citet{sta03} show that comptonization of such a starlight radiation\footnote{that dominates over the energy densities of the other photon fields in the jet comoving frame, in particular over the energy density of the synchrotron photons \citep{sta03};} within the kpc-scale jet (its brightest knot A) by the synchrotron-emitting electrons in the equipartition magnetic field can possibly account for the TeV emission detected by {\it HEGRA} Cherenkov Telescope from the direction of that source \citep{aha03}. However, subsequent observations of M~87 by {\it Whipple} Telescope \citep{leb04} gave only upper limits for its emission at the $0.4 - 4$ TeV photon energy range, suggesting, although not strictly implying, variability of the VHE $\gamma$-ray signal. Such variability, clearly confirmed by the most recent {\it HESS} observations which established the presence of a variable (on the time scale of years) VHE $\gamma$-ray source within 0.1 deg ($\sim 30$ kpc) of the M~87 central region \citep{bei04}, questions the possibility that the extended kpc-scale jet is responsible for the $3 - 4 \sigma$ detections of M~87 by {\it HEGRA} and {\it HESS} telescopes. On the other hand, the upper limits imposed in this way put interesting constraints on the magnetic field within the M~87 large-scale jet, an issue which is of general importance in astrophysics \citep[see, e.g., recent monograph on the cosmic magnetic fields by][]{val04}, and in particular in the physics of extragalactic jets \citep{dey02}.

Here we comment in more details on the high energy $\gamma$-ray emission of the M~87 kpc-scale jet, resulting from the IC scattering on the stellar photon field. We take into account a relativistic bulk velocity of the emitting region as well as Klein-Nishina regime of the electron-photon interaction. We emphasize an important aspect of the presented model: IC scattering on the starlight emission by the synchrotron-emitting electrons is \emph{inevitable}, and involves neither the unknown target photon field, nor the additional unconstrained source of the ultrarelativistic high-energy electrons. In particular, following our previous approach presented in \citet{sta03}, we `reconstruct' the electron energy distribution from the \emph{known} broad-band synchrotron spectrum of a given jet region, and then estimate the IC flux for the \emph{known} target photon field. Therefore, our discussion is independent of any model of particle acceleration. This constitutes an important difference with the other models proposed in the literature in the context of the VHE $\gamma$-ray emission of M~87 system \citep{pfr03,rei04}.

The paper is organized as follows. In section 2 we present the formalism used in order to evaluate the high-energy $\gamma$-ray emission of knot A in the M~87 jet. In section 3 we compare the estimated fluxes with the observations reported in the literature. In section 4 we discuss implications of the obtained lower limit on the magnetic field to the theoretical models of FR I structures and M~87 jet in particular. General conclusions are presented in the last section 5.

\section{The Model}

M~87 is a giant radio galaxy at the distance of $16$ Mpc (leading to the conversion scale $78$ pc/$''$), possessing a famous 2-kpc-long (projected) one-sided jet bright in radio, optical and X-rays due to its \emph{synchrotron} emission. Here we evaluate the IC radiation of the brightest (in the radio and optical regimes) knot in this jet, knot A, placed at the projected distance $12.4''$ from the active galactic center. In particular, we evaluate the IC flux from this knot due to comptonization of the stellar light of the host galaxy by the synchrotron-emitting electrons, resulting in a high-energy $\gamma$-ray emission. Below we assume that the jet flow is relativistic at the position of knot A, as justified by the one-sidedness of the entire jet structure.

\subsection{Electron energy distribution}

Detailed broad-band (radio-to-X-ray) observations of the M~87 large-scale jet, reported by, e.g., \citet{owe89,bir91,mei96,spa96,per01,mar02,wil02}, circumscribe well the synchrotron spectrum of knot A. Hereafter we refer to the analysis by \citet{wil02}, which indicates that the energy distribution of the synchrotron-emitting electrons can be approximated by a broken power-law,
\begin{equation}
n'_{\rm e}(\gamma) = K'_{\rm e} \times \left\{\begin{array}{ccc} \gamma^{-p} & {\rm for} & \gamma \leq \gamma_{\rm br} \\ \gamma_{\rm br}^{q} \, \gamma^{-(p + q)} & {\rm for} & \gamma > \gamma_{\rm br} \end{array} \right. \quad ,
\end{equation}
\noindent
with the spectral indices $p = 2.3$ and $q = 1.6$. Here $n'_{\rm e} \equiv \int n'_{\rm e}(\gamma) \, d \gamma$ is the comoving number density of the electrons\footnote{We follow the notation with the primed quantities measured in the jet comoving frame and the bare ones if given in the observer rest frame, with exception of the magnetic field intensity $B$ as well as electron Lorentz factors $\gamma$, which always refer to the emitting plasma rest frame.} contributing to the observed emission of knot A. The electron break Lorentz factor
\begin{equation}
\gamma_{\rm br} = \left({4 \pi \, m_{\rm e} c \, \nu_{\rm br} \over e \, \delta \, B} \right)^{1/2} \approx 2.7 \times 10^6 \, \delta^{-0.5} \, B_{-4}^{-0.5} \quad ,
\end{equation}
\noindent
that corresponds to the observed synchrotron break frequency $\nu_{\rm br} = 10^{15}$ Hz for the emitting plasma magnetic field $B \equiv B_{-4} \, 10^{-4}$G and Doppler factor $\delta$. The cut-off energies $\gamma_{\rm min}$ and $\gamma_{\rm max}$ are basically unconstrained, but the synchrotron origin of the X-ray jet emission indicates $\gamma_{\rm max} / \gamma_{\rm br} \geq 10 - 100$. We normalize the number of electrons which contribute to the observed synchrotron luminosity of knot A,
\begin{equation}
[\nu L_{\nu}]_{\rm syn} = 4 \pi \, \delta^4 \, V' \, [\nu' j'_{\nu'}]_{\rm syn} \quad ,
\end{equation}
\noindent
where $V'=V_{obs} / \delta$ is the volume filled by those particles which are `seen' at the given moment, and the synchrotron emissivity times the frequency as measured in the jet rest frame, $\nu' j'_{\nu'}$, is
\begin{equation}
[\nu' j'_{\nu'}]_{\rm syn} = {c \sigma_T \over 48 \pi^2} \, B^2 \, [\gamma^3 n'_e(\gamma)]_{\gamma = (4 \pi \, m_{\rm e} c \, \nu' / e \, B)^{1/2}} \quad ,
\end{equation}
\noindent
where $\nu' = \nu / \delta$ \citep[see, e.g.,][]{sta03}. Hence, for a given observed synchrotron break luminosity $L_{\rm br} = 3 \times 10^{41}$ erg s$^{-1}$ \citep{wil02} and other parameters as discussed above, one obtains the product that will be used later $V' K'_{\rm e} = 1.8 \times 10^{60} \, \delta^{-3.65} \, B_{-4}^{-1.65}$. Note, that the above expressions apply either if knot A is a moving blob or a stationary shock \citep[see a discussion in][]{sik97}.

\subsection{Starlight photon field}

For the target starlight photons at the position of knot A we assume roughly isotropic distribution in the galactic rest frame and strongly anisotropic in the jet rest frame, due to the relativistic jet velocity. We take the characteristic observed frequency of the optical-NIR bump due to the elliptical host of M~87 as $\nu_{\rm star} = 10^{14}$ Hz \citep[see, e.g.,][]{mul04}. We also evaluate the appropriate starlight energy density at the position of knot A \emph{directly} from the observations of M~87 host galaxy. 

\citet{you78} showed that the distribution of stars in M~87 host galaxy agrees with the King model for the distances $>1'' - 2''$ from the active center, while at the smaller scales an additional population of massive stars is present due to a supermassive black hole perturbing central region of the galaxy \citep[see also][]{mac97}. Let us, therefore, consider the starlight of the `unperturbed' population of the evolved stars, for which the emissivity can be approximated by
\begin{equation}
j_{\rm star}(r) = j_0 \, \left[ 1 + \left({r \over r_{\rm c}}\right)^2\right]^{-3/2} \quad {\rm for} \quad r < r_{\rm t} \quad ,
\end{equation}
\noindent
where $r$ is the radius as measured from the galactic center, $r_{\rm c}$ is the core radius for the galaxy, and $r_{\rm t}$ is the appropriate tidal radius. We normalize this distribution to the luminosity density profile $\rho_{\rm L}(r) = 4 \pi \, j_{\rm star}(r)$ in the $I$ band, as given by \citet{lau92}. For the parameters $\rho_{\rm L}(1 \, {\rm kpc}) = (3 - 4) \times L_{\odot}$ pc$^{-3}$, $r_{\rm c} = 0.55$ kpc and $r_{\rm t} = 68$ kpc \citep{lau92,you78} we obtain $j_0 \approx (3 - 4) \times 10^{-22}$ erg s$^{-1}$ cm$^{-3}$. The intensity of the starlight emission in a given direction specified by the azimuthal angle $\zeta \equiv \cos^{-1} \kappa$, can be found by integrating the stellar emissivity along a ray,
\begin{equation}
I_{\rm star}(r, \kappa) = \int_0^{l_{\rm max}} j_{\rm star}\left( \sqrt{r^2 + l^2 + 2 r l \kappa} \right) \, dl \quad , 
\end{equation}
\noindent
where the outer boundary of the host galaxy is
\begin{equation}
l_{\rm max} = - r \kappa + \sqrt{r_{\rm t}^2 - r^2 + r^2 \kappa^2} \quad ,
\end{equation}
\noindent
as discussed in \citet{tsa95}. By integrating further over the solid angle, one can find the required value for the starlight energy density at a given position $r$ from the core,
\begin{equation}
U_{\rm star}(r) = {2 \pi \over c} \, \int_{-1}^{+1} I_{\rm star}(r, \kappa) \, d\kappa
\end{equation}
\noindent
\citep[see equation 26 in][]{tsa95}. With all the parameters as discussed in this section, we obtain the $I$-band stellar energy density $U_{\rm star}(1 \, {\rm kpc}) \approx 10^{-10}$ erg cm$^{-3}$.

There are several reasons why the above estimate should be considered as a safe lower limit. First, the obtained value refers to the $I$-band energy density, and not to the bolometric one. We also do not include any effects of photon absorption by gas or dust. Second, in the analysis above, contribution from the additional population of massive stars from the central cusp was neglected. Also, the 2-kpc-long M~87 jet is surrounded by the filaments of the optically emitting cluster gas \citep[e.g.,][]{spa04}, which additionally contributes to the intensity of the optical radiation around knot A. Finally, we note that synchrotron emission produced within knot A -- peaked in the jet rest frame at similar frequencies to the stellar emission, although characterized by much broader energy distribution and much lower energy density than the starlight photon field -- may be important for the IC emission of the highest energy electrons, increasing total IC flux at highest photon energies. 

\subsection{Inverse-Compton emission}

The high-energy emissivity of knot A due to IC scattering on monoenergetic and mono-directional (in the jet rest frame) starlight photon field, including proper relativistic effects in the Klein-Nishina regime, can be found from the approximate expression given by \citet[equation 20 therein]{aha81} as
\begin{equation}
[\nu' {j'}_{\nu'}]_{\rm ic} = {c \, \sigma_{\rm T} \over 4 \pi} \, U_{\rm star} \, \nu_{\rm star}^{-2} \, {\nu'}^2 \, \int_{\gamma_0}^{\gamma_{\rm max}} \, {n'_{\rm e}(\gamma) \over \gamma^2} \, f(\epsilon', \epsilon'_{\rm star}, \gamma, \mu') \, d\gamma \quad .
\end{equation}
\noindent
Here $\epsilon' \equiv h \, \nu' / m_{\rm e} c^2$, $\epsilon'_{\rm star} \equiv h \, \nu'_{\rm star} / m_{\rm e} c^2$, $\theta' \equiv \cos^{-1} \mu'$ is the scattering angle, and
\begin{equation}
f(\epsilon', \epsilon'_{\rm star}, \gamma, \mu') = 1 + {{w'}^2 \over 2(1-w')} - {2w' \over v'(1-w')} + {2{w'}^2 \over {v'}^2(1-{w'})^2}
\end{equation}
\noindent
where $v' = 2(1-\mu') \, \epsilon'_{\rm star} \gamma$ and $w' = \epsilon' / \gamma$. The lower limit of the integral over $\gamma$, given by the condition $f \ge 0$, is
\begin{equation}
\gamma_0 = {\epsilon' \over 2} \left[ 1+\left(1+{2 \over (1-\mu') \epsilon' \epsilon'_{\rm star}}\right)^{1/2} \right] \quad .
\end{equation}
\noindent
Hence, using the well known relativistic transformations $\epsilon_{\rm star}' = \epsilon_{\rm star} \Gamma$ (where $\Gamma$ is a jet bulk Lorentz factor), $\epsilon' = \epsilon / \delta$, and $\mu' = (\mu - \beta)/(1-\beta \mu)$, one can find the observed IC flux as
\begin{equation}
[\nu S_{\nu}]_{\rm ic} = {1 \over d_{\rm L}^2} \, \delta^4 \, V' \, [\nu' {j'}_{\nu'}]_{\rm ic} \quad .
\end{equation}
\noindent
Note, that the IC fluxes evaluated below do not depend on the poorly known volume of the emission region, as normalization of the electron energy distribution to the observed synchrotron emission allows us to find the product $V' K'_{\rm e}$, which is then inserted into eq. 12. 

The above expressions for the IC emissivity are correct only if $\epsilon' / \epsilon'_{\rm star} \gg 1$ and $\gamma \gg 1$, as considered in this paper. For a general case the appropriate formula for $[\nu' {j'}_{\nu'}]_{\rm ic}$ is much more complicated \citep[see][]{aha81,bru00}.

\subsection{Kinematic factors}

High jet-counterjet brightness asymmetry for the 2-kpc jet structure in M~87 ($> 450$ in optical energy range), as well as other morphological properties of this object, led \citet{bic96} to conclude that the appropriate bulk Lorentz factor at the position of knot A is $\Gamma \sim 3 - 5$, and the jet viewing angle $\theta \sim 30^0 - 35^0$. \citet{hei97} considered similar values for $\Gamma$ but also discussed smaller jet inclinations, $\theta \sim 20^0$. For such a choice of $\Gamma$ and $\theta$ one gets the jet Doppler factor $\delta = [\Gamma (1 - \beta \cos \theta)]^{-1} \sim 1 - 3$. We note, that superluminal velocities detected by {\it Hubble} Space Telescope downstream of knot HST-1 in M~87 jet \citep[distances $0.''8 - 10''$ from the center;][]{bir99}, as well as the X-ray and optical month-to-year variability of the HST-1 knot emission \citep{har03,per03}, suggest even higher values for the jet Doppler factor, albeit characterizing the jet flow upstream with respect to knot A. One should be aware that the high Doppler factor ($\delta > 3$) of the kpc-scale jet in M~87 would require a small jet inclination, leading in turn to a decrease of the energy density of the starlight photons at the position of knot A, as the physical distance of this region from the core increases with the decreasing jet viewing angle. These effects introduce, however, only minor changes of the evaluated magnetic field intensity, as it depends rather weakly on the exact value of $U_{\rm star}$ (for which we take in this paper a very safe lower limit, anyway). We note that according to the discussion in \citet{zav02}, a large jet viewing angle ($\theta > 20^0$) in M~87 radio galaxy is consistent with a lack of any polarized radio emission from its core explained in terms of obscuration of the active nucleus by a dense, multi-phase nuclear disk depolarizing the core emission.

\section{The Results}

Spectral energy distribution of the high-energy $\gamma$-ray IC emission of knot A is presented on figures 1 and 2 for a different magnetic field intensity, two jet viewing angles $\theta = 30^0$, $20^0$ and the bulk Lorentz factors $\Gamma = 3$, $5$. The Thomson part of this emission extends up to photon energies of the order of $10^{10} - 10^{11}$ eV. Below we compare the expected IC fluxes for different photon energies to the upper limits imposed by the observations of {\it EGRET} observatory and ground-based Cherenkov Telescopes: {\it HESS}, {\it HEGRA} and {\it Whipple}.

\subsection{{\it EGRET} observations}

{\it EGRET} observations of Virgo cluster imply the photon flux $F(> 100 \, {\rm MeV}) < 2.2 \times 10^{-8}$ cm$^{-2}$ s$^{-1}$ \citep{rei03}. When converted to the energy flux density assuming power-law emission with the spectral index $\alpha_{\gamma} = 0.65$, as expected in the Thomson regime of the IC emission of knot A, this reads as $\nu S_{\nu}(100 \, {\rm MeV}) < 2.3 \times 10^{-12}$ erg cm$^{-2}$ s$^{-1}$. 

Figures 3 and 4 show the expected flux of knot A at the observed $100$ MeV photon energy as a function of the magnetic field intensity, for the jet viewing angles $\theta = 30^0$ and $20^0$, and the bulk Lorentz factors $\Gamma = 3$ and $5$. The vertical lines indicate the appropriate equipartition value that can be found from the synchrotron spectrum of the knot A, 
\begin{equation}
B_{\rm eq} = 330 \, \delta^{- 5/7} \, \mu{\rm G} \quad ,
\end{equation}
\noindent
as given by \citet{kat04}. Note that the adopted value of $B_{\rm eq}$ refers to the energy equipartition between the jet magnetic field and ultrarelativistic electrons, with possible contribution from the non-radiating particles neglected. The {\it EGRET} observations thus imply $B > 30 - 100$ $\mu$G for any choice of the kinematic factors considered here, corresponding roughly to $B / B_{\rm eq} > 0.1 - 0.5$. This constraint does not necessarily mean a departure from the magnetic field--radiating particles energy equipartition but, interestingly enough, already excludes a class of models involving a very weak jet magnetic field. Note in this context, that models postulating `loss-free channel' within the jet, i.e. a reservoir of high-energy electrons residing in the jet regions where the magnetic field is extremely low enabling thus the particles to travel along the jet without radiative energy losses, are also excluded. That is because of the dramatic radiative losses suffered by such high-energy electrons when propagating through the intense stellar radiation field of the host galaxy. Such a channel in the context of M~87 jet was discussed by \citet{owe89}.

\subsection{{\it Whipple} observations}

{\it Whipple} observations give the $99 \%$ CL upper limit to the VHE $\gamma$-ray photon flux of M~87 radio galaxy $F(> 0.4 \, {\rm TeV}) < 6.9 \times 10^{-12}$ cm$^{-2}$ s$^{-1}$ \citep{leb04}. As can be seen in figures 1 and 2, such an energetic emission from the jet would be produced entirely in the Klein-Nishima regime. 

Figures 5 and 6 show the expected photon flux of knot A at the observed photon energies $h \nu_0 > 0.4$ TeV, where
\begin{equation}
F(> h \nu_0) = \int_{\nu_0} {[\nu S_{\nu}]_{\rm ic} \over h \nu^2} \, d\nu \quad ,
\end{equation}
\noindent
(see equation 12) as a function of the magnetic field intensity, for the jet viewing angles $\theta = 30^0$ and $20^0$ considered here and the bulk Lorentz factors $\Gamma = 3$ and $5$. The vertical lines indicate again the appropriate equipartition magnetic field. One can see that, independently of the {\it EGRET} constraints, the {\it Whipple} observations imply magnetic field intensity within the discussed jet region $B > 50 - 60$ $\mu$G, or in other words $B / B_{\rm eq} > 0.2 - 0.4$.

\subsection{{\it HEGRA} and {\it HESS} observations}

{\it HEGRA} observations resulted in a $4 \sigma$ detection of high-energy $\gamma$-ray flux from the direction of M~87 with the photon flux $F(> 0.73 \, {\rm TeV}) \approx 0.96 \times 10^{-12}$ cm$^{-2}$ s$^{-1}$ \citep{aha03}. Recent {\it HESS} observations resulted also in marginal detection of the M~87 system at the $3 - 4 \sigma$ level, however with the photon fluxes $F(> 0.73 \, {\rm TeV}) \approx 0.4 \times 10^{-12}$ cm$^{-2}$ s$^{-1}$ in 2003, and $F(> 0.73 \, {\rm TeV}) \approx 0.15 \times 10^{-12}$ cm$^{-2}$ s$^{-1}$ in 2004 \citep{bei04}. This clearly indicates a variability of the high-energy $\gamma$-ray emission of this source, and therefore gives the \emph{upper limits} for the VHE radiation of knot A.

Figures 7 and 8 show the expected photon flux of knot A at the observed photon energies $ h \nu_0 > 0.73$ TeV as a function of the magnetic field intensity, for the jet viewing angles $\theta = 30^0$ and $20^0$, and the bulk Lorentz factors $\Gamma = 3$ and $5$. The vertical lines indicate the appropriate equipartition magnetic field. The most recent {\it HESS} observations imply therefore the magnetic field within knot A as strong as $B > 300$ $\mu$G, i.e. $B / B_{\rm eq} > 1 - 2$, again for any choice of the kinematic factors considered here. Let us emphasize here once again, that in evaluating the IC fluxes the safe \emph{lower limit} on the starlight energy density at the position of knot A was considered. Therefore, in face of the {\it HESS} observations one can firmly conclude that a weak subequipartition jet magnetic field is excluded in M~87. 

\section{Discussion}

Our study indicates that the magnetic field within the brightest knot A of the M~87 jet, placed at $\sim 1-3$ kpc from the active nucleus, is $B \gtrsim 300 \, \mu{\rm G} \gtrsim B_{\rm eq}$ (if the jet viewing angle is in the range $\theta = 20^0 - 30^0$ and the jet bulk Lorentz factors $\Gamma = 3 - 5$). On the other hand the upper limit to the magnetic field intensity within knot A can be found from the upper limit imposed on the magnetic field energy flux, $L_{\rm B} \equiv {1 \over 8} R^2 c \Gamma^2 B^2 \leq L_{\rm j}$, where $L_{\rm j} \sim {\rm few} \times 10^{44}$ erg s$^{-1}$ is the total power of M~87 jet \citep{owe00} and $R \approx 60$ pc is the radius of radio structure at the position of knot A. This gives roughly $B_{\rm max} < 1000$ $\mu$G. Now let us discuss a few issues related to the above derived magnetic field lower limit.

\subsection{Synchrotron Continuum}

The low-energy electrons within knot A, if present, can also inverse-Compton up-scatter the starlight photons to the observed X-ray photon energies. For example, in order to produce $1$ keV emission in this process, one has to involve electrons with Lorentz factors $\gamma \sim 50 / \delta$. The spectrum of such low-energy electrons is unknown, because their synchrotron radio emission ($\nu < 10$ MHz) cannot be directly observed. If, however, electron energy distribution of the form given in equation 1 can be indeed extrapolated down to $10$ MeV electron energies, the IC energy flux density $[\nu S_{\nu}]_{\rm ic} \propto \nu^{1-0.65}$ at $h \nu = 1$ keV is expected to be $\approx 10^{-15} - 10^{-14}$ erg cm$^{-2}$ s$^{-1}$ for all the parameters as discussed above. This is more than an order of magnitude below the energy flux detected by {\it Chandra}, $\nu S_{\nu}(1 \, {\rm keV}) = 3.4 \times 10^{-13}$ erg cm$^{-2}$ s$^{-1}$ \citep{mar02}. We note that the variability of the X-ray emission of knot A on the time scale of years \citep{har97} indicates that the X-rays detected from this region are synchrotron in origin.

\citet{wil02} reported X-ray spectral indices characterizing different parts of the M~87 jet that are significantly flatter than the appropriate optical-to-X-ray power-law slopes, although still relatively steep, $\alpha_{\rm X} > 1$ \citep[but see][]{wil04}. If real in the case of knot A, such spectral flattenings would be difficult to explain as resulting from the comptonization of the starlight emission by the low-energy electrons (i.e., as an evidence for the transition from the synchrotron to the IC spectral component), according to the discussion above. Hence, they could indicate most probably pile-up effects occurring at the high-energy tail of the electron energy distribution emitting synchrotron X-rays. \citet{der02} showed that similar pile-up features can appear in the synchrotron spectra of extragalactic large-scale jets due to a decrease of the IC cooling rate of the high-energy electrons in the Klein-Nishina regime. In particular, \citet{der02} considered cosmic microwave background radiation as the seed photon field for the IC emission, obtaining spectral hardenings of the synchrotron jet continua at the observed X-ray photon energies for highly relativistic jet. In the case considered here, it is the starlight radiation which dominates the photon energy density within the emitting region ($U_{\rm star} / U_{\rm CMB} \gg 100$), and hence potential pile-up effects should be pronounced at lower energies of the emitted synchrotron photons. This already indicates that the X-ray spectral flattenings in the M~87 jet -- if real -- must be produced by another physical mechanism, such as for example stochastic particle acceleration processes acting within extended turbulent regions of the jet flows, as discussed by \citet{sta02}. We note here that \citet{mei96} reported a `marginal but significant' flattening of the optical spectra at the boundaries of some parts of M~87 jet. Such flatteninigs could result from an efficient acceleration of the high-energy electrons at the jet edges, independently from the X-ray spectra formed closer to the jet central regions.

\subsection{Particle Acceleration}

\citet[who considered synchrotron and adiabatic energy losses of radiating electrons]{hei97} showed that a long extension of the optical jet in the M~87 radio galaxy (as compared with the propagation length of the synchrotron emitting electrons) can be explained without invoking continuous particle acceleration only if the jet is relativistic ($\Gamma = 3 - 5$) and the jet magnetic field on kpc-scale is below equipartition ($B \leq 0.2 - 0.7 \, B_{\rm eq}$). As discussed in this paper, the later condition is not likely to be fulfilled, and hence some kind of particle re-acceleration acting continuously within the jet is needed. As pointed out by many authors \citep[e.g.,][]{dey86}, boundary shear layers of the large-scale jets are very likely to be highly turbulent, and hence are very likely to be the sites for the efficient second-order Fermi acceleration of the jet particles.\footnote{We invoke this well physically and mathematically described process. However it stands here for a number of different mechanisms, which are expected to accelerate particles within the jet boundary layer. One can mention in this context reconnection processes in highly perturbed and sheared magnetic fields or small scale, local oblique/weak shocks formed in regions of forced supersonic turbulence.} As discussed further by \citet{sta02}, the maximum energies the electrons can reach in such a process can be very high ($>$ TeV), and the resulting electron energy distribution can deviate from a simple power-law behavior at the highest energy range, reflecting conditions within the jet flows. In this scenario, limb-brightenings of the jet are expected. It is therefore interesting to note, that the limb-brightenings are indeed observed in some of the FR I jets in radio \citep[M~87,][]{owe89}, optical \citep[3C~66B and 3C~264;][respectively]{mac91,cra93} and X-rays \citep[Centaurus~A;][]{hard03}. In the case of the M~87 jet, however, the optical structure is narrower than the radio structure \citep{spa96}, displaying in addition different polarization pattern when compared to the radio one \citep[which is most likely determined by the boundary shear layer morphology,][]{per99}. This fact does not necessarily imply that the optically emitting electrons cannot be accelerated predominantly at the jet boundaries, as the turbulent mixing of the jet matter connected with the entrainment processes -- important in the case of FR I jets -- can play a role in this context \citep[see][]{dey02}. The presence of the turbulent mixing layer between the jet and the surrounding medium was also suggested to explain Faraday rotation measures for the small-scale jet in M~87 source \citep{zav03}. However, further investigation of this problem is impeded due to hardly known jet internal structure.

\subsection{Jet Magnetic Field}

The comoving energy density of the magnetic field within knot A is limited roughly by $U'_{\rm B} = B^2 / 8 \pi \geq 3.6 \times 10^{-9}$ erg cm$^{-3}$. As discussed above, this value is most likely higher than the energy density of the radiating electrons. This may indicate the energy equipartition between the jet magnetic field and relativistic protons present within knot A, $U'_{\rm B} \sim U'_{\rm rel, \, p} \gtrsim U'_{\rm rel, \, e}$, although rough lower limits obtained in this paper do not enable for further quantitative analysis of this issue.

{\it VLBI} measurements often allow one to infer magnetic field intensity from the low-frequency spectral break in the radio emission of the (sub) pc-scale jet modeled in terms of synchrotron self-absorption process. In the case of the M~87 jet, this method gives $B_{\rm VLBI} < 0.2$ G at $R_{\rm VLBI} \ll 0.06$ pc \citep{rey96}. If the magnetic energy flux in a jet is constant, then one should expect magnetic field intensity at the position of knot A to be roughly $B = (\Gamma_{\rm VLBI}/ \Gamma) \, (R_{\rm VLBI} / R) \, B_{\rm VLBI}\ll 300$ $\mu$G, where we put $\Gamma_{\rm VLBI} / \Gamma \approx 2$ and the jet radius at the position of the considered knot $R \approx 60$ pc \citep{owe00}. Hence one can conclude that some kind of amplification of the jet magnetic field has to take place between the parsec and kiloparsec scales, although all the estimates presented above should be taken with caution, as some arbitrary assumptions on the jet magnetic field structure were invoked \citep[but see][]{hug04}.

The suggested amplification of the jet magnetic field can take place at the extended shock wave located within knot A \citep[see][]{med99}. However, another (in some respect more `natural' in this object) possibility is offered by an action of the turbulent dynamo processes discussed in this context by \citet{dey80} \citep[see also, e.g.,][]{gva88,urp02}. That is because the Kelvin-Helmholtz instabilities occurring inevitably at the edges of the jet are supposed to create large-scale eddies at the flow boundaries, which amplify the jet magnetic field and develop a highly turbulent mixing layer between the jet and the surrounding medium. Such turbulent shear layers play a crucial role in mass entrainment (and so deceleration) of the FR I outflows \citep{bic94}, and in acceleration of the jet particles, influencing also polarization properties of the jets \citep{lai80}. In a framework of this model, in order to allow for the turbulent dynamo process to proceed at all, the M~87 jet has to be dynamically dominated by the (cold) particles on the small scales. On the larger scales the mass entrainment process decelerates the flow gradually, slowly amplifying the jet magnetic field (to the value exceeding at some point the energy equipartition with the radiating electrons) and creating a turbulent boundary layer that spreads into the jet interior. This process accomplishes at $\sim 1 - 2$ kpc from the active center (knot A and beyond), where the jet magnetic field reaches an approximate equipartition with the energy density of the jet particles, while the jet flow itself becomes fully turbulent, disappearing into the outer amorphous radio lobe \citep[see low-frequency radio maps of M~87 by][]{owe00}. Again, further discussion of this issue is hindered due to unknown details of the turbulent dynamo process and the jet internal structure.

\section{Conclusions}

Here we discuss how the present upper limits on the high-energy $\gamma$-ray emission of the kpc-scale jet in the M~87 radio galaxy can be used to estimate magnetic field strength in the brightest knot of the jet. We obtain a `safe' lower limit $B > 300$ $\mu$G, which indicates a very likely departure from the minimum power condition in a sense that the magnetic field energy density within knot A is higher than the energy density of the radiating ultrarelativistic electrons. We speculate that the high magnetic field in knot A of the M~87 jet is due to turbulent dynamo processes related to interaction of the jet with the surrounding medium. 

\acknowledgments

\L .S. was supported by the grant 1-P03D-011-26 and partly by the Chandra grants G02-3148A and G0-09280.01-A. A.S. was supported by the NASA contract NAS8-39073 and Chandra Avard G02-3148A. M.O. and M.S. were supported by the grant PBZ-KBN-054/P03/2001. \L .S. acknowledges also very useful comments from F. Aharonian, K. Aldcroft, C.C. Cheung, D.S. De Young, D.E. Harris, and J. Kataoka.

{}

\begin{figure}
\includegraphics[scale=1.50]{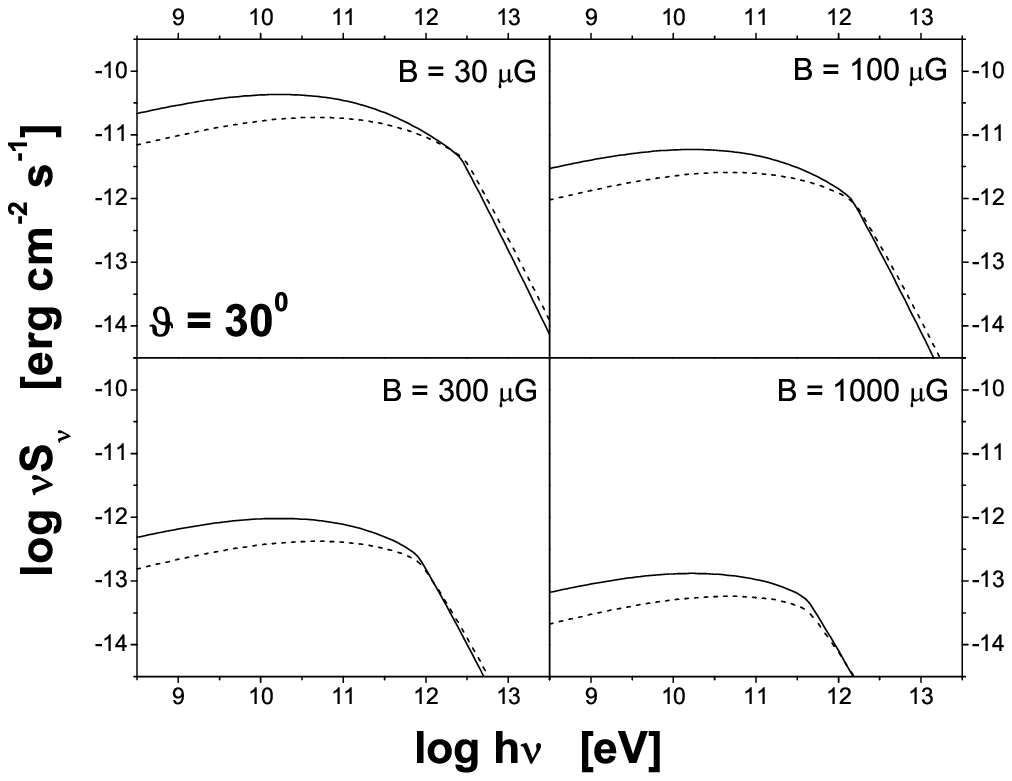}
\caption{Spectral energy distribution of high-energy $\gamma$-ray inverse-Compton emission of knot A for the jet magnetic field $B = 0.3 \times 10^{-4}$ G, $10^{-4}$ G, $3 \times 10^{-4}$ G, $B = 10^{-3}$ G (as indicated at each panel), the jet viewing angle $\theta = 30^0$ and the jet bulk Lorenz factors $\Gamma = 5$ and $3$ (solid and dashed lines, respectively).}
\end{figure}

\begin{figure}
\includegraphics[scale=1.50]{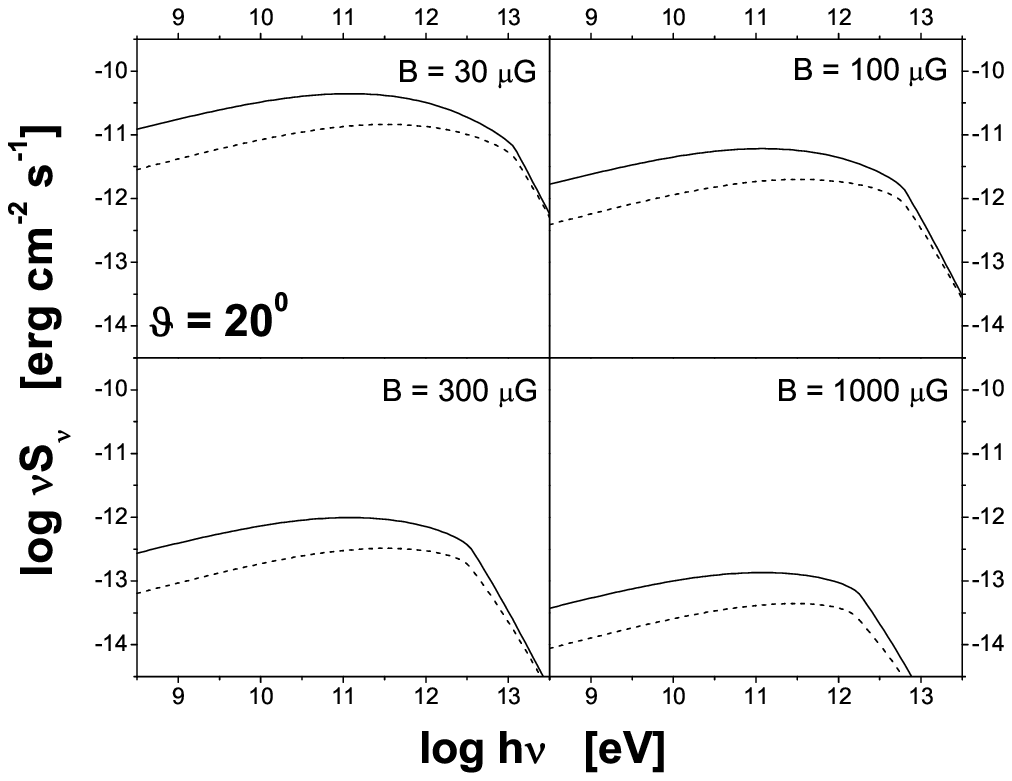}
\caption{Spectral energy distribution of high-energy $\gamma$-ray inverse-Compton emission of knot A for the jet magnetic field $B = 0.3 \times 10^{-4}$ G, $10^{-4}$ G, $3 \times 10^{-4}$ G, $B = 10^{-3}$ G (as indicated at each panel), the jet viewing angle $\theta = 20^0$ and the jet bulk Lorenz factors $\Gamma = 5$ and $3$ (solid and dashed lines, respectively).}
\end{figure}

\begin{figure}
\includegraphics[scale=1.50]{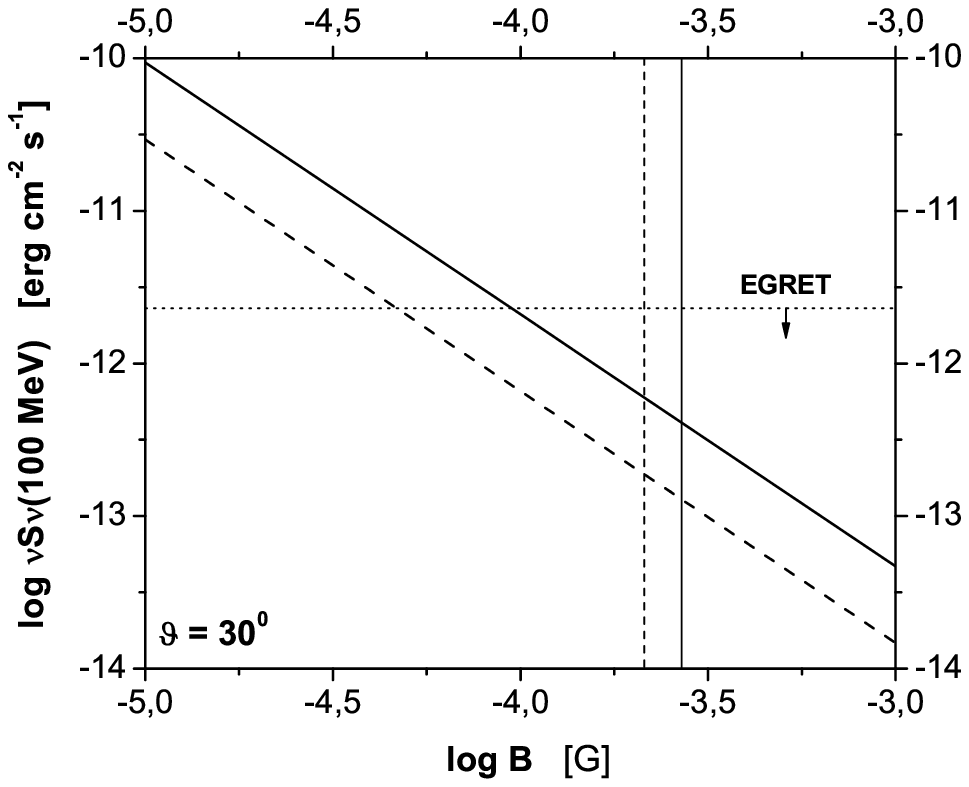}
\caption{Constraints on the jet magnetic field within knot A imposed by the {\it EGRET} observations (dotted horizontal line), for the jet viewing angle $\theta = 30^0$ and the jet bulk Lorenz factors $\Gamma = 5$ and $3$ (solid and dashed lines, respectively). Vertical lines denote the equipartition magnetic field for $\Gamma = 5$ and $3$ (solid and dashed lines, respectively).}
\end{figure}

\begin{figure}
\includegraphics[scale=1.50]{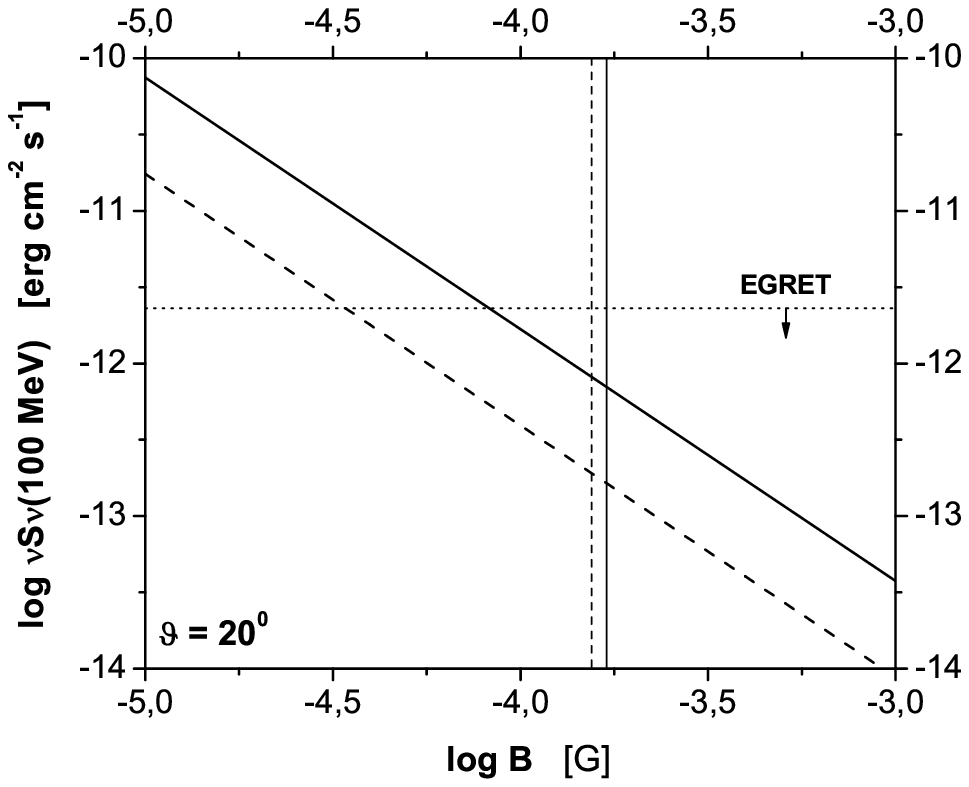}
\caption{Constraints on the jet magnetic field within knot A imposed by the {\it EGRET} observations (dotted horizontal line), for the jet viewing angle $\theta = 20^0$ and the jet bulk Lorenz factors $\Gamma = 5$ and $3$ (solid and dashed lines, respectively). Vertical lines denote the equipartition magnetic field for $\Gamma = 5$ and $3$ (solid and dashed lines, respectively).}
\end{figure}

\begin{figure}
\includegraphics[scale=1.50]{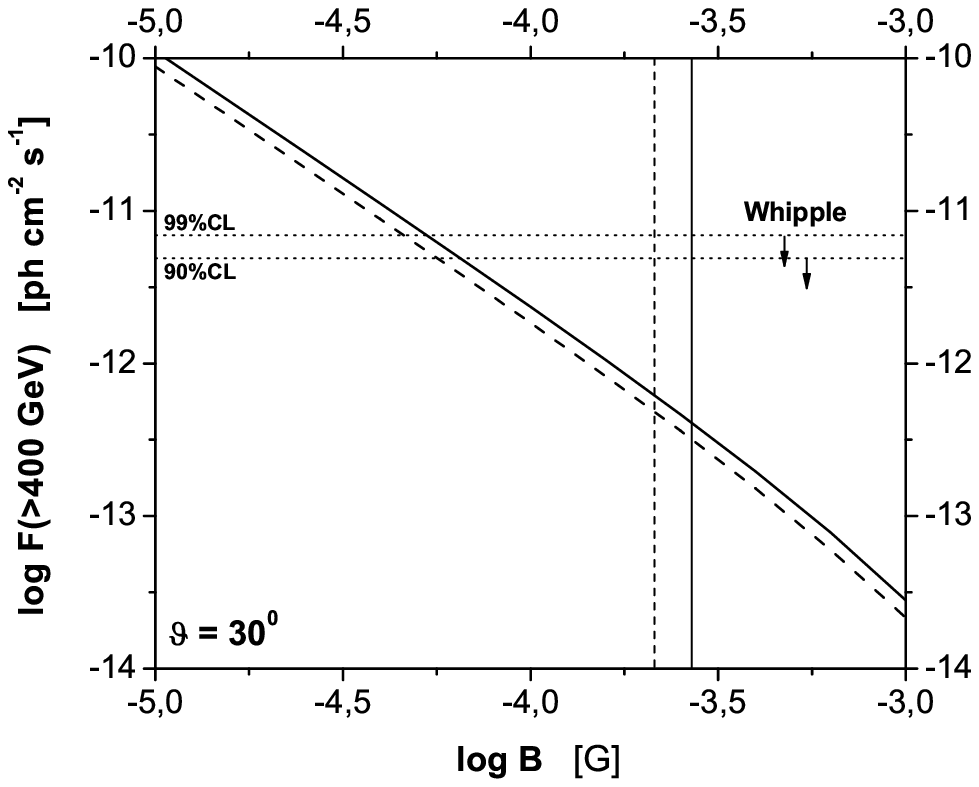}
\caption{Constraints on the jet magnetic field within knot A imposed by the {\it Whipple} observations (dotted horizontal lines), for the jet viewing angle $\theta = 30^0$ and the jet bulk Lorenz factors $\Gamma = 5$ and $3$ (solid and dashed lines, respectively). Vertical lines denote the equipartition magnetic field for $\Gamma = 5$ and $3$ (solid and dashed lines, respectively).}
\end{figure}

\begin{figure}
\includegraphics[scale=1.50]{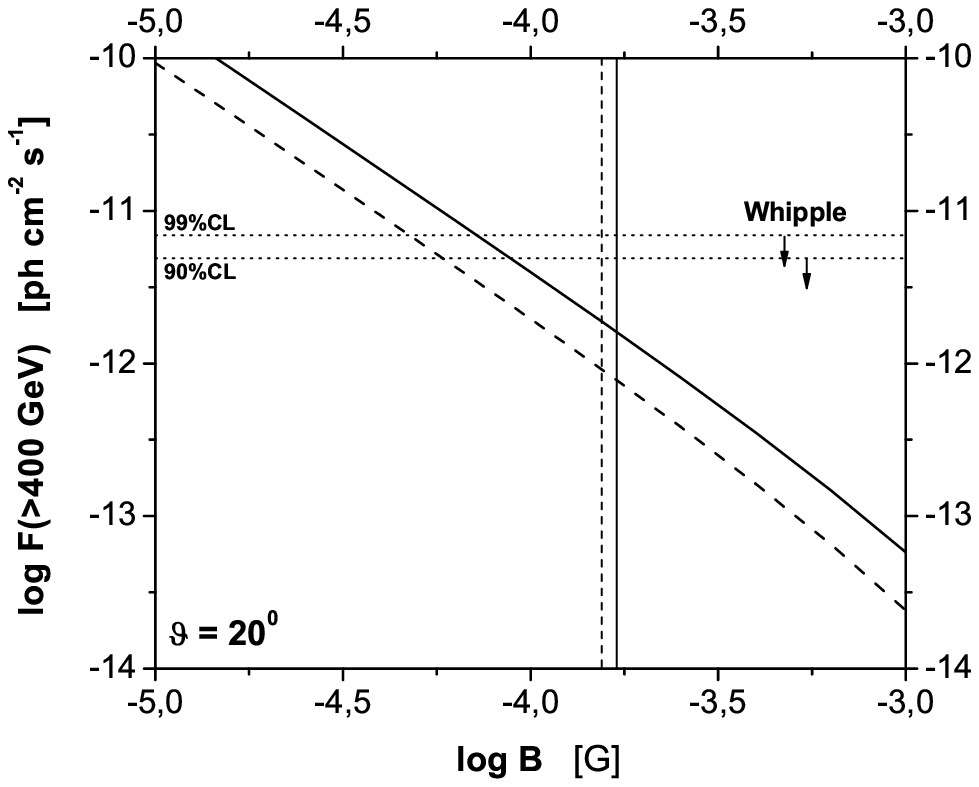}
\caption{Constraints on the jet magnetic field within knot A imposed by the {\it Whipple} observations (dotted horizontal lines), for the jet viewing angle $\theta = 20^0$ and the jet bulk Lorenz factors $\Gamma = 5$ and $3$ (solid and dashed lines, respectively). Vertical lines denote the equipartition magnetic field for $\Gamma = 5$ and $3$ (solid and dashed lines, respectively).}
\end{figure}

\begin{figure}
\includegraphics[scale=1.50]{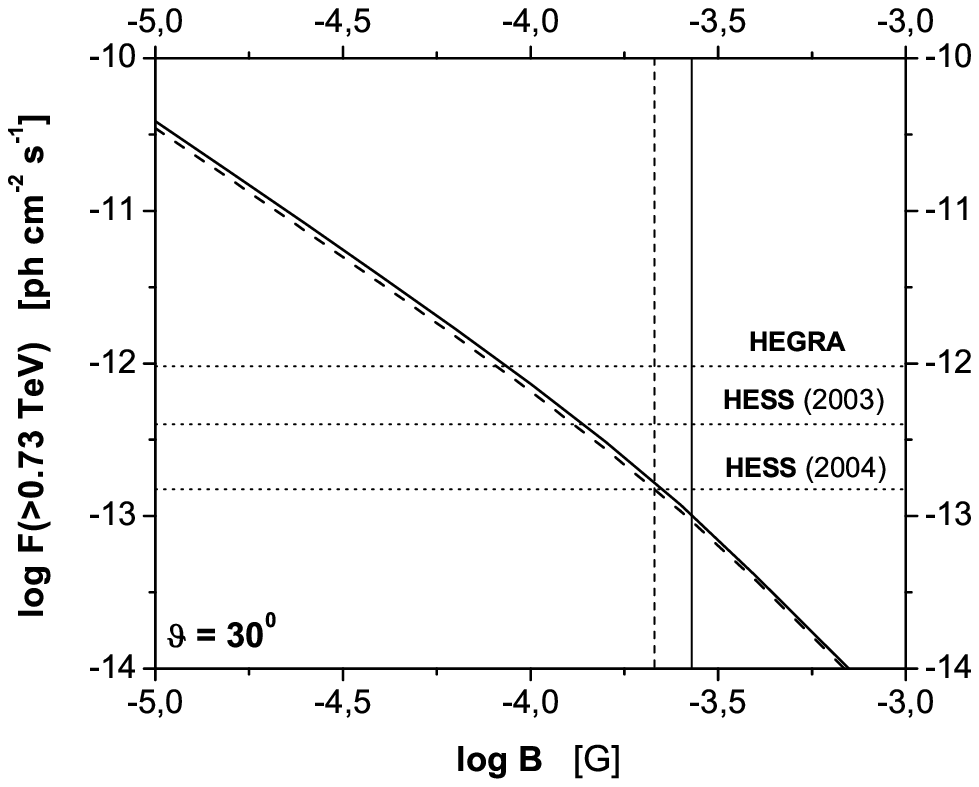}
\caption{Constraints on the jet magnetic field within knot A imposed by the {\it HEGRA} and {\it HESS} observations (dotted horizontal lines), for the jet viewing angle $\theta = 30^0$ and the jet bulk Lorenz factors $\Gamma = 5$ and $3$ (solid and dashed lines, respectively). Vertical lines denote the equipartition magnetic field for $\Gamma = 5$ and $3$ (solid and dashed lines, respectively).}
\end{figure}

\begin{figure}
\includegraphics[scale=1.50]{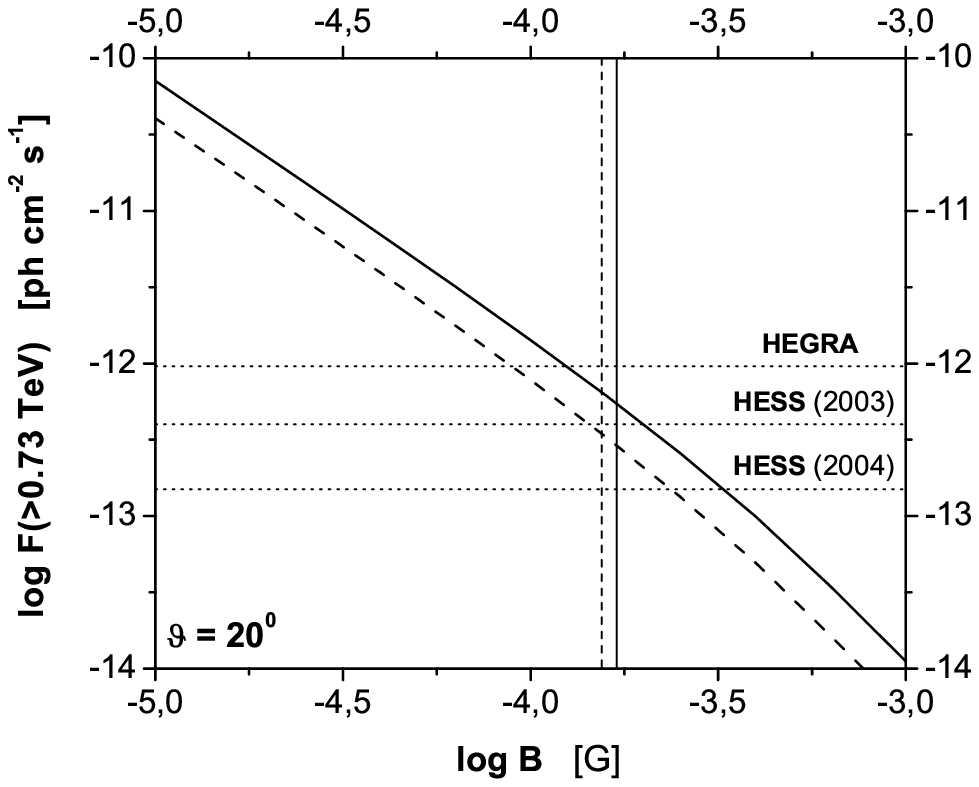}
\caption{Constraints on the jet magnetic field within knot A imposed by the {\it HEGRA} and {\it HESS} observations (dotted horizontal lines), for the jet viewing angle $\theta = 20^0$ and the jet bulk Lorenz factors $\Gamma = 5$ and $3$ (solid and dashed lines, respectively). Vertical lines denote the equipartition magnetic field for $\Gamma = 5$ and $3$ (solid and dashed lines, respectively).}
\end{figure}

\end{document}